# Overview of the LAMOST survey in the first decade

Hongliang Yan,[1,2] Haining Li,[1] Song Wang,[1] Weikai Zong,[3] Haibo Yuan,[3] Maosheng Xiang,[4] Yang Huang,[6] Jiwei Xie,[6,7] Subo Dong,[8] Hailong Yuan,[1] Shaolan Bi,[3] Yaoquan Chu,[9] Xiangqun Cui,[10,11] Licai Deng,[1] Jianning Fu,[3] Zhanwen Han,[12] Jinliang Hou,[2,13] Guoping Li,[10,11] Chao Liu,[2,14] Jifeng Liu,[1,2,15] Xiaowei Liu,[5] Ali Luo,[1,2] Jianrong Shi,[1,2] Xuebing Wu,[8,16] Haotong Zhang,[1] Gang Zhao,[1,2] and Yongheng Zhao[1,2,*]

*Correspondence: yzhao@nao.cas.cn

## GRAPHICAL ABSTRACT

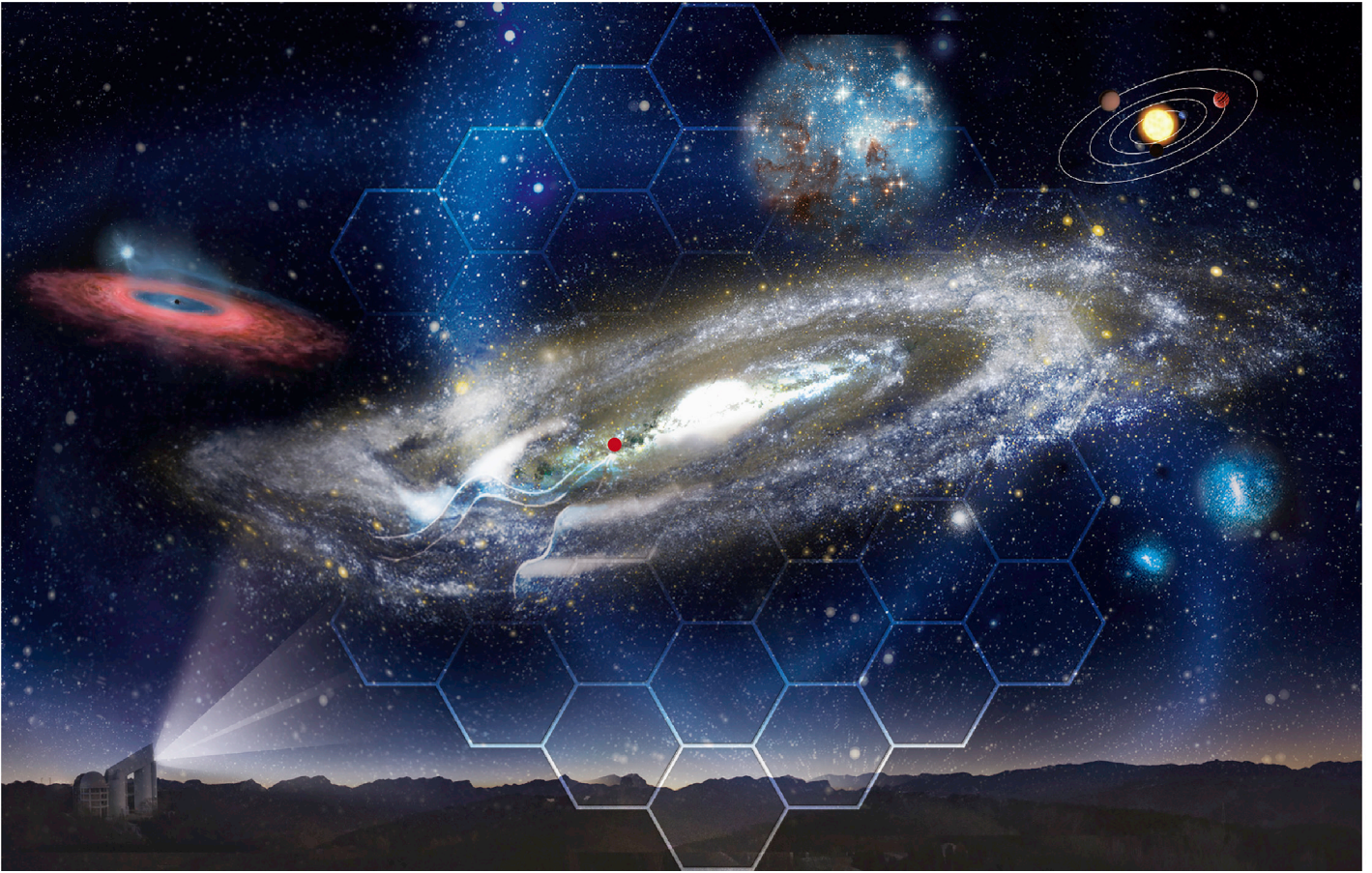

## PUBLIC SUMMARY

- LAMOST is an innovative telescope designed with both a large-aperture and a wide-FOV for astronomical spectroscopic survey

- LAMOST observed over 10 million objects in our Galaxy, and constructed the largest spectroscopic dataset

- LAMOST data changed the astrophysical viewpoint in the fields including stars, the Milky Way, exoplanets, and black holes

# Overview of the LAMOST survey in the first decade

Hongliang Yan,[1,2] Haining Li,[1] Song Wang,[1] Weikai Zong,[3] Haibo Yuan,[3] Maosheng Xiang,[4] Yang Huang,[5] Jiwei Xie,[6,7] Subo Dong,[8] Hailong Yuan,[1] Shaolan Bi,[3] Yaoquan Chu,[9] Xiangqun Cui,[10,11] Licai Deng,[1] Jianning Fu,[3] Zhanwen Han,[12] Jinliang Hou,[2,13] Guoping Li,[10,11] Chao Liu,[2,14] Jifeng Liu,[1,2,15] Xiaowei Liu,[5] Ali Luo,[1] Jianrong Shi,[1,2] Xuebing Wu,[8,16] Haotong Zhang,[1] Gang Zhao,[1,2] and Yongheng Zhao[1,2,*]

[1]CAS Key Laboratory of Optical Astronomy, National Astronomical Observatories, Beijing 100101, China
[2]School of Astronomy and Space Science, University of Chinese Academy of Sciences, Beijing 100049, China
[3]Department of Astronomy, Beijing Normal University, Beijing 100875, China
[4]Max-Planck Institute for Astronomy, Königstuhl, 69117, Heidelberg, Germany
[5]South-Western Institute for Astronomy Research, Yunnan University, Kunming 650500, China
[6]School of Astronomy and Space Science, Nanjing University, Nanjing 210093, China
[7]Key Laboratory of Modern Astronomy and Astrophysics, Ministry of Education, Nanjing University, Nanjing 210093, China
[8]Kavli Institute for Astronomy and Astrophysics, Peking University, Beijing 100871, China
[9]University of Science and Technology of China, Hefei 230026, China
[10]National Astronomical Observatories/Nanjing Institute of Astronomical Optics & Technology, Chinese Academy of Sciences, Nanjing 210042, China
[11]CAS Key Laboratory of Astronomical Optics & Technology, Nanjing Institute of Astronomical Optics & Technology, Nanjing 210042, China
[12]Yunnan Observatories, Chinese Academy of Sciences, Kunming 650011, China
[13]Key Laboratory for Research in Galaxies and Cosmology, Shanghai Astronomical Observatory, Chinese Academy of Sciences, Shanghai 200030, China
[14]Key Laboratory of Space Astronomy and Technology, National Astronomical Observatories, CAS, Beijing 100101, China
[15]WHU-NAOC Joint Center for Astronomy, Wuhan University, Wuhan 430072, China
[16]Department of Astronomy, School of Physics, Peking University, Beijing 100871, China
*Correspondence: yzhao@nao.cas.cn

The Large Sky Area Multi-Object Fiber Spectroscopic Telescope (LAMOST), also known as the Guoshoujing Telescope, is a major national scientific facility for astronomical research located in Xinglong, China. Beginning with a pilot survey in 2011, LAMOST has been surveying the night sky for more than 10 years. The LAMOST survey covers various objects in the Universe, from normal stars to peculiar ones, from the Milky Way to other galaxies, and from stellar black holes and their companions to quasars that ignite ancient galaxies. Until the latest data release 8, the LAMOST survey has released spectra for more than 10 million stars, ∼220,000 galaxies, and ∼71,000 quasars. With this largest celestial spectra database ever constructed, LAMOST has helped astronomers to deepen their understanding of the Universe, especially for our Milky Way galaxy and the millions of stars within it. In this article, we briefly review the characteristics, observations, and scientific achievements of LAMOST. In particular, we show how astrophysical knowledge about the Milky Way has been improved by LAMOST data.

## INTRODUCTION

A direct and effective way to study the Universe is to observe celestial bodies (also called celestial objects) using telescopes. To obtain statistical information about these objects and study the physical rules that govern them, observations are usually performed in such a way that the telescope does not have a specific target or a position to point at; rather, it scans the entirety of the observable sky within several years. This is called a survey.

Surveys are a powerful way to gain knowledge about the Universe; however, they require the telescope to have a large aperture and a wide field of view (FOV) at the same time, which is difficult to design due to the limitation of optical systems.[1] A large aperture ensures that the telescope can capture as many photons as possible in a unit of time, thereby achieving a lower detection limit and improving data quality, whereas a wide FOV allows the telescope to capture as large an area as possible in a single exposure, thus improving the efficiency of the survey.

The innovative design of the Large Sky Area Multi-Object Fiber Spectroscopic Telescope (LAMOST) (Figure 1) was proposed in the 1990s by Chinese astronomers.[2] To fulfill the aforementioned requirements, LAMOST was specifically designed as an active reflecting Schmidt telescope with the capacity for continuously changing its mirror surface to achieve a series of different reflecting Schmidt systems. LAMOST consists of three major components (Figure 1): an active aspherical correcting mirror (Ma), a spherical primary mirror (Mb), and a focal surface with fibers. These three major components are lined up along the Earth's longitude (the meridian plane). Ma is located in the north, Mb in the south, and the focal surface is between them. They form a +25° inclination to the horizontal from Ma to Mb. Both mirrors are segmented and Ma can change its surface shape, a property continuously controlled by the active optics technique. LAMOST has a variable effective aperture from 3.6 to 4.9 m and a FOV of 5° (see supplemental information and Table S1 for more details).

To date, LAMOST has released ∼17 million spectra, and this number continues to increase. Since 2014, the LAMOST dataset has been the largest spectral dataset ever obtained. These data have enabled research in a number of cutting-edge topics in astronomy, especially stellar physics and Milky Way (MW) sciences. In this paper, we review the scientific achievements obtained through the LAMOST survey. In the following section, we present information about the LAMOST survey. After that, we review its scientific achievements and show how our understanding of the Universe has changed. A brief perspective is then presented, with a short summary to conclude the article.

## LAMOST SURVEY

### Observation

The LAMOST survey typically consists of 9–10 months of observing nights during its 1-year cycle. Observations usually begin in September or October and end in June of the following year. After the observing season, the telescope can take a break in July and August for instrument maintenance. The observational strategies are discussed and scheduled by the scientific committee, which consists of scientists from a broad range of astrophysical fields to ensure that all of the scientific cases of interest are covered by the LAMOST survey.[3]

To improve the instrumental performance, LAMOST initiated the pilot survey in October 2011, which lasted for 9 months. After the pilot survey, LAMOST launched its regular survey from September 2012, which lasted for 5 years. The 1-year pilot survey plus the 5-year regular survey is usually referred to as the low-resolution spectroscopic survey (LRS), as the spectra obtained during this period were at a resolution power of $R \equiv \lambda / \Delta\lambda \sim 1,800$. The LAMOST-LRS consisted of two main parts.[3] The LAMOST Experiment for Galactic Understanding and Exploration (LEGUE) survey[4] aimed to study the stars and the MW itself. The LAMOST Extra Galactic Survey (LEGAS) focused mainly on the sciences of galaxies and cosmology. There are also several highly specialized programs, such as the LAMOST Spectroscopic Surveys for Galactic Anti-Center[5,6] (LSS-GAC) or the LAMOST-Kepler project.[7,8] For the LRS, the spectral wavelength coverage was from 370 to 900 nm, and the magnitude range of the targets was $9.0 \leq r_{mag} \leq 17.5$ mag.

Following a full year of testing, a new survey mode with a resolution power of $R \sim 7,500$ began in September 2018. This new mode was referred to as the medium-resolution spectroscopic survey[9] (MRS). It was planned that the LA-MOST-MRS would also last for 5 years, until June 2023. In the MRS mode, the spectral wavelength coverage is from 495 to 535 nm for the blue band and from 630 to 680 nm for the red band. The magnitude range of targets for the MRS is $9.0 \leq G_{mag} \leq 15.0$ mag. It is noted that approximately half of the observing time was for acquiring MRS spectra since 2018, while the other half was for the LRS mode to cover unobserved areas from the LAMOST-LRS or newly added targets of interest. LAMOST can readily switch between these two modes (see the supplemental information and Table S2 for more details regarding the LRS and MRS).

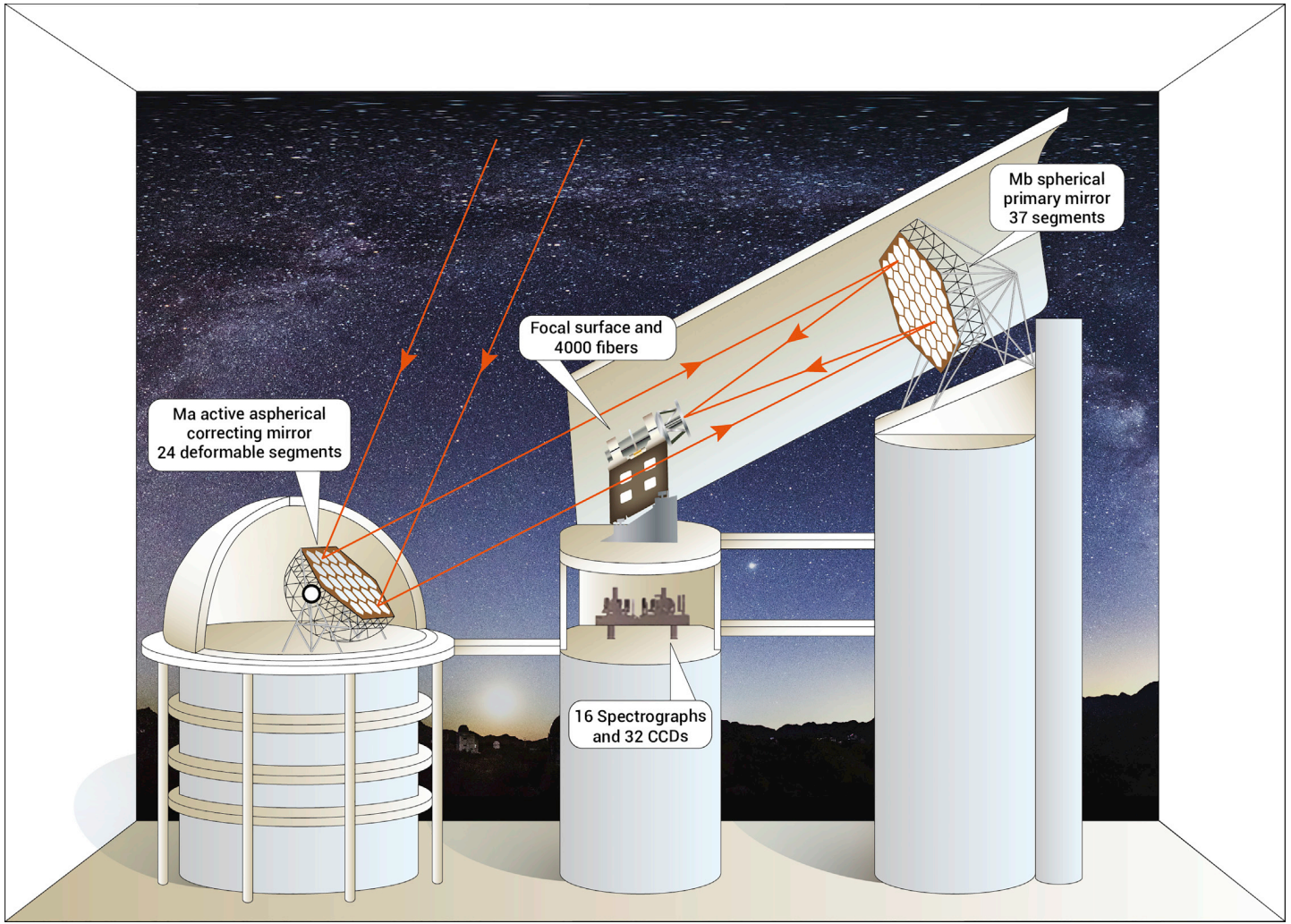

**Figure 1. Structure and optical path of LAMOST** The characteristic appearance of LAMOST is evidently different from telescopes using the traditional optical design. The red solid line with arrows indicate the path of light traveling from the objects to the focal surface. CCD, Charge-Coupled Device

In Figure 2, we show representative spectra for observed objects (panel A) and the observation footprint (panels B and C) of LAMOST. The coverage of the sky is essentially homogeneous, and the gap between ∼285°–315° of right ascension corresponds to the summer maintenance period. For more detailed information about the survey, readers are referred to Zhao et al. (2012)[3] and Liu et al. (2020)[9] for the LRS and MRS, respectively.

### Data release

LAMOST has released ∼17 million spectra to the public before the end of 2021 and has released data every year. In general, the data obtained in an observation season will be released in March of the next year to Chinese users, then to international users after 18 months. The dataset released to the public contains not only new data from the previous observation season but also older data for all of the past seasons that have been reduced and analyzed by the most up-to-date version of the data pipeline (see supplemental information for details), the LAMOST Stellar Parameter Pipeline[10] (LASP). Information for the latest data release 8 (DR8) is shown for the LRS and MRS in Tables S3 and S4, respectively.

Aside from the LASP, a number of research teams have developed their own pipelines, such as the LAMOST Stellar Parameter Pipeline at Peking University[11] (LSP3) and SPAce.[12] For more detailed information on the data pipeline, the readers are referred to Luo et al. (2015).[10]

In Figure 3, we show the overall information derived from LASP for stars in the LRS. Stars observed by LAMOST are dominated by F, G, and K types in either the main sequence (the longest and most stable phase in the lifetime of a star during which hydrogen fusion occurs in the core) or evolved phases (after a star has exhausted the hydrogen in its core).

### Comparison with other spectroscopic surveys

A number of other spectroscopic surveys have overlapped in time with LA-MOST. For ease of comparison, we divide these surveys into two groups. "Current surveys" consist of survey projects that either have been completed or are still ongoing, while "imminent surveys" consist of surveys that have just started or are planned in the near future, with data products to be released in the early 2020s.

In Table 1, we briefly list the features of these representative survey projects together with those of the LAMOST survey. Among the current surveys,[1,13–17] LA-MOST,[1] the Sloan Extension for Galactic Understanding and Exploration (SEGUE),[14] and the Apache Point Observatory Galactic Evolution Experiment (APOGEE)[15] mainly cover the northern sky, while the Radial Velocity Experiment (RAVE),[13] Galactic Archaeology High Efficiency and Resolution Multi-Element Spectrograph (GALAH),[16] and the Gaia-European Southern Observatory (Gaia-ESO)[17] cover the southern sky. The most significant advantage of the LAMOST survey is its uniquely high efficiency in obtaining stellar spectra. The survey products of LAMOST significantly enlarge the current database, providing an important reservoir of stellar spectra in the northern sky, especially in the direction of the anti-Galactic center. However, the LAMOST survey also forms a good complement with the APOGEE, GALAH, and Gaia-ESO surveys in wavelength coverage and sky area. For example, in the current surveys group, the APOGEE survey uniquely provides infrared spectra, which allows astronomers to study the Galactic bulge, while the GALAH survey covers the southern sky, supplementing various Galactic components and structures that cannot be observed from the northern hemisphere.

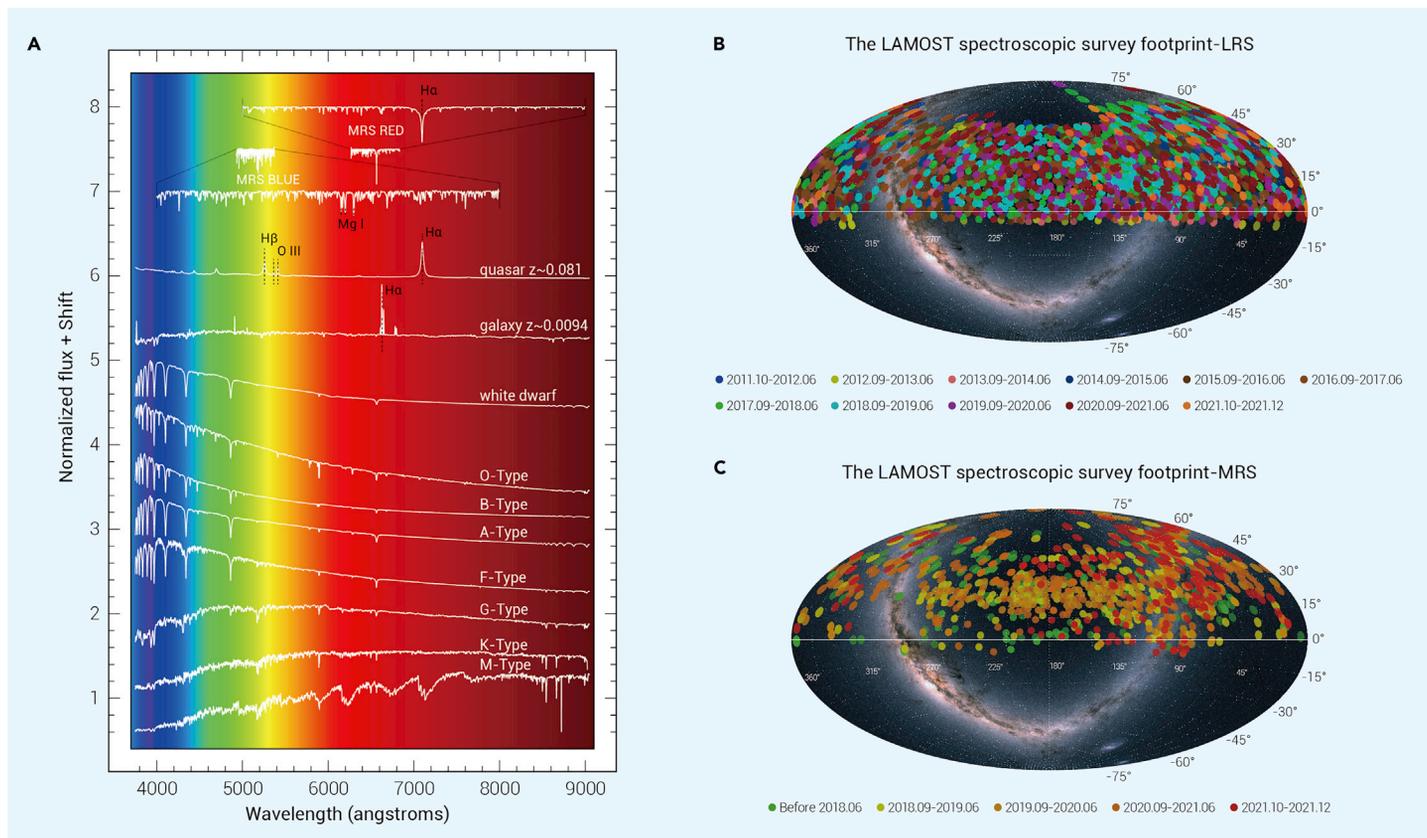

**Figure 2. LAMOST spectra and observation footprint** Representative spectra for observed objects (A) and the observation footprint (B and C). At the top of (A), we show the blue and red bands of the MRS spectra for a G-type star, followed by the LRS spectra of a quasar (QSO), an emission-line galaxy, a white dwarf (WD), and 7 stars with spectral types O, B, A, F, G, K, and M, respectively. In (B) and (C), we show the footprints of the LRS and MRS, respectively. The dotted lines in (B) and (C) indicate the equatorial coordinate system.

Although the imminent surveys[18–22] are overall technically superior to those of the current ones, there is nonetheless still great potential for synergy with LAMOST. The imminent surveys will provide spectra at higher resolutions for more distant stars. The specific areas of focus will be the Galactic bulge (Sloan Digital Sky Survey-V [SDSS-V][19]), thick disk (SDSS-V and the William Herschel Telescope Enhanced Area Velocity Explorer [WEAVE][20]), halo (Dark Energy Spectroscopic Instrument [DESI][15]), nearby galaxies (WEAVE), and very distant galaxies (DESI). Compared with the northern imminent surveys, the current LAMOST data, even without consideration of future updates, provides a competitively large number of spectra for stars in the thin disk and solar vicinity with a similar level of data volume.

Aside from spectroscopic surveys, LAMOST also synergizes with astrometric/photometric surveys and spaceborne missions, which is described using scientific examples in the following section.

## SCIENTIFIC ACHIEVEMENTS OF THE LAMOST SURVEY

LAMOST opened a new window for astrophysicists to rediscover the Universe and MW by providing a vast collection of spectra that was previously unachievable. Since its commissioning phase, LAMOST data have produced a number of scientific results.[23–28] In Figure 4, we show statistical information from the Web of Science (https://www.webofscience.com/) about scientific publications based on LAMOST data. Up to the end of 2021, more than 900 scientific papers had been published in academic journals worldwide. Both the number of papers and citations have been increasing steadily. LAMOST data also gained broad attention from the international astronomical community, with approximately 40% of LAMOST scientific papers published by research teams abroad. In this section, we briefly review the research and scientific achievements accomplished using LAMOST data.

### Obtain fundamental parameters and value-added catalogs

Revealing the physics of stars, planets, and the MW requires accurate characterization of stellar physical properties. In particular, unraveling the structure and assembly history of the MW with LAMOST has been established through obtaining accurate knowledge about the physical information delivered from the spectra for millions of stars, such as their radial velocities (RVs), masses, ages, and chemical compositions. A number of works have demonstrated that precise stellar atmospheric parameters can be estimated from the $R \sim 1,800$ spectra, with a precision of 100 K in effective temperature ($T_{eff}$), 0.1 dex in surface gravity ($\log g$), 0.1 dex in metallicity ([Fe/H]) given a spectral signal-to-noise ratio (S/N) higher than 50 per pixel.[29–31] Furthermore, for the first time, individual elemental abundances for 16 elements (C, N, O, Na, Mg, Al, Si, Ca, Ti, Cr, Mn, Fe, Co, Ni, Cu, and Ba) have been delivered from the exceptionally large number of LRS spectra.[30,32] These achievements have been built on both the innovative spectral modeling[30,32,33] and the effective synergy between LAMOST and other surveys, especially the high-resolution spectroscopic surveys such as APOGEE[15] and GALAH,[16] the photometric transit survey of the Kepler[34] mission, and the astrometric survey of the Gaia[35] mission. These synergies provide uniform and high-precision training and calibration sets, which provides an important basis for LAMOST stellar parameter estimation.[29–31,36–38]

The stellar age is of vital importance when disentangling stellar populations and tracing the Galactic formation and evolution history. Limited by the availability of precise stellar atmospheric parameters, robust stellar age estimates were restricted to only a small number of stars (~10,000) in the solar neighborhood.[39,40] Precise stellar parameters obtained through LAMOST spectra thus allowed robust age estimates for millions of stars beyond the solar neighborhood[41–45] (Figure 5), providing cornerstone data for unraveling the assembly and evolution history of the MW[46–51] and studying stellar astrophysics.[52–54]

Complementary to the official DR, extensive value-added stellar parameter catalogs created by LAMOST users[6,12,30,41,44,55–57] are publicly available. For example, value-added catalogs for the LSS-GAC[5,6] provide RVs, $T_{eff}$, $logg$, [Fe/H], [$\alpha$/Fe], [C/H], and [N/H] derived using LSP3,[11,58] as well as the extinction, distance, and orbital parameters further inferred from these data.

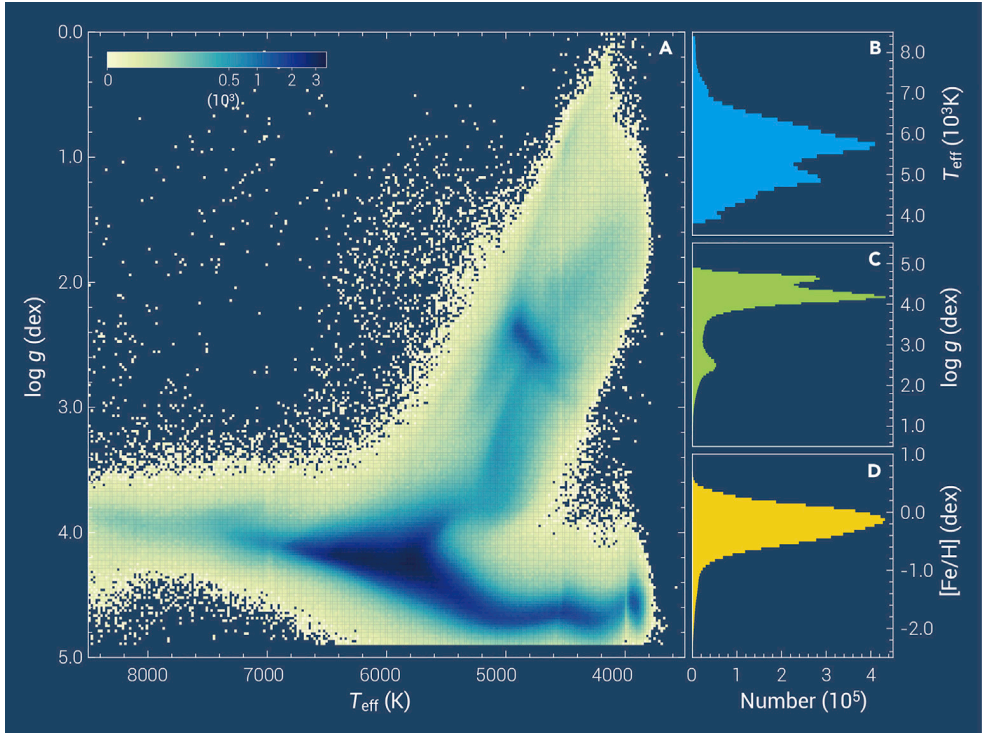

**Figure 3. Overall distribution of the LAMOST-LRS stars released by DR8 version 1.0** (A) The Hertzsprung-Russell diagram of the entire LAMOST-LRS sample with different number densities coded by different colors (indicated at top left). (B–D) The distribution of these stars in single dimension of stellar parameters, including effective temperature (blue), surface gravity (green), and metallicity (yellow), respectively.

### Gain deep insights into the stars

**Stars with chemical peculiarity.** Some stars (usually not in great numbers) show peculiar chemical features—for example, an overabundance of certain elements. Such stars are important since they either reflect formation scenarios different from the classical theory or they may have experienced a distinct evolution history compared to the majority of stars in the MW.

One type of peculiar star is the lithium (Li)-rich low-mass evolved star or the so-called Li-rich giant.[59] Approximately 1% of evolved stars show anomalously high Li abundances ($A_{Li}$) that conflict with the standard stellar evolutionary model.[60] The number of Li-rich stars has been dramatically expanded dozens of times using LAMOST data.[61–66] LAMOST has discovered the most Li-rich giant stars (Figure 6) in the MW, with an $A_{Li}$ of 4.51 dex.[67] The Li content in this star is ∼3,000 times higher than that of the Sun. LAMOST has also found a metal-poor turnoff star with extremely high Li abundance.[68] Li-rich giants were previously thought to be mainly red giant branch (RGB) stars; however, such viewpoints have been revolutionized with the help of LAMOST data.[69] With the exception of highly evolved stars, ∼80% of Li-rich giants are at the red clump (RC) phase, whereas RGBs are in the minority.[62,69] Detailed analyses reveal a series of new features in Li-rich giants,[69] indicating different origins of these two types. One possible scenario for the overabundant Li in RC stars was proposed based on the GALAH and LAMOST surveys. The helium (He) flash that ignites the He core of stars may be the same process that enriches Li, indicating that Li production in low-mass evolved stars could be a universal phenomenon.[70–72]

The LAMOST survey has enabled systematic searching for stars with peculiar chemical features and greatly expanded the known sample, including carbon stars,[73–75] nitrogen-enriched stars,[76,77] $\alpha$-deficient stars,[78] stars enriched in the elements produced by the s- and r-processes,[52,79] and magnetic chemically peculiar stars.[80] The significantly enlarged samples provide unique opportunities for astronomers to investigate the origins of such objects, often leading to the discovery of new stellar phenomena, previously unknown stellar evolution pathways, and new scientific laws.

**Low-mass and massive stars.** M dwarf stars are the most abundant objects in the MW. They are perfect tracers of the chemodynamic properties of the solar neighborhood and the most preferred candidates around which to search for exoplanets.[81] In addition, being among the most luminous objects in the Universe, M giants are regarded as ideal tracers to explore the structure of the outer disk and distant halo of the MW. A number of efforts have been made to search for M dwarf[82–84] and M giant[85] stars using LAMOST data since its pilot survey, which has also resulted in a better understanding of the physical properties of M-type stars.[86] Recently, Xiang et al.[87] made a further advancement by presenting the largest catalog of M-type stars using LAMOST DR5. In total, 39,796 M giants and 501,152 M dwarfs have been identified. Until the latest DR8, the number of M-type stars released has been increased to more than 700,000. Such a database provides a valuable source for the study of Galactic star formation histories and evolution through the window of M-type stars.

Massive stars (eg, O/B-type stars) are rare, especially in comparison to M-type stars, but are equally important. Through powerful stellar winds and supernova (SN) explosions, massive stars can strongly influence the chemical and dynamic evolution of galaxies. Important progress in the field of O/B stars has also been made based on LAMOST data. Using LAMOST DR5, the largest O/B-type star catalog of 16,032 such stars has been established,[88] for which key physical parameters have also been derived.[89] A number of peculiar O-type stars were simultaneously discovered, such as the very rare Oe stars, which present O-type spectra with the emission of H lines but without N III λ4634−4640−4642 or He II λ4686 emission features.[90]

**Stars with intrinsic brightness variation.** LAMOST offers an opportunity to support spaceborne photometry. A series of projects[91–93] have been performed to combine LAMOST spectra and Kepler photometry, with the expectation of fruitful advances in stellar physics.

Stellar flares and spots, activities that encompass a range of phenomena produced by dynamo action in the interior of a star, are related to stellar magnetism, rotation, differential rotation, and subphotospheric convection. Improving our knowledge of their origin and the prediction of their strength is vital to understanding stellar processes within our Solar System. If a solar superflare were to occur, the survival of the Earth's atmosphere and human life would be jeopardized. The physical processes of flares and spots can be investigated by monitoring the luminosity variations of stars[94] as well as by measuring the strength of particular lines—for instance, Ca II K or H$\alpha$, using LAMOST spectra.[95] Statistical analyses of a large sample of stars suggest that 10% of superflare events occurring on other stars show magnetic strengths similar to, or even weaker than, solar values.[96] Using the dedicated measurements of photometric variability for solar-like stars, another study finds that the Sun is less active than its identical cousins.[97,98]

Pulsating stars characterized by luminosity variations are found across almost the entire Hertzsprung-Russell diagram; their brightness variations provide a unique opportunity to probe their interiors and chemical profiles via asteroseismology.[99] Seismic characterization shapes stellar evolutionary theory by requiring the development of more detailed and precise theories on stellar processes.[100] As an input constraint, atmospheric parameters derived from spectra help to reduce the searching space before proceeding with seismic modeling[101] or mass estimation of companion stars detected using the pulsation timing method.[102] New methods to search for pulsating stars are advanced through well-derived atmospheric parameters such as $logg$ and $T_{eff}$ provided by LAMOST. For instance, the instability stripe of the pulsating H atmosphere white dwarfs is pure—in other words, a star evolving through that region in the Hertzsprung-Russell diagram must be pulsating.[103] The LAMOST-observed white dwarfs[104] and hot B subdwarfs[105,106] have been used to discover pulsating white dwarfs[107] and to provide extensive high-quality spectra that compliment photometric observations from ongoing or upcoming spaceborne missions.[108,109]

**Table 1.** Large-scale spectroscopic survey projects

| Survey name | R | Wavelength coverage (μm) | No. fibers | Limiting magnitude (mag) | No. spectra (10⁶) | Sky coverage (mainly) | Schedule |
|---|---|---|---|---|---|---|---|
| Current survey | | | | | | | |
| LAMOST[a] | 1,800 | 0.37–0.90 | 4,000 | r ≤ 17.8 | 11.0 | Northern sky | 2011– |
| | 7,500 | 0.49–0.54 | | G ≤ 15 | 6.0 | | 2018– |
| | | 0.60–0.68 | | | | | |
| RAVE[b] | 7,500 | 0.84–0.88 | 400 | I < 12 | 0.5 | Southern sky | 2003–2013 |
| SEGUE[c] | 2,000 | 0.38–0.92 | 320 | g < 19 | 0.4 | Northern sky | 2004–2009 |
| APOGEE[d] | 22,500 | 1.51–1.69 | 300 | H < 12.2 | 0.6 | Northern sky | 2011–2020 |
| GALAH[e] | 28,000 | 0.47–0.49 | 400 | G ≤ 13 | 0.8 | Southern sky | 2015– |
| | | 0.56–0.59 | | | | | |
| | | 0.64–0.68 | | | | | |
| | | 0.75–0.79 | | | | | |
| Gaia-ESO[f] | 17,000 | 0.40–0.68 | 132 | V ≤ 19 | 0.1 | Southern sky | 2013-2018 |
| | 47,000 | | 8 | V ≤ 16.5 | | | |
| Imminent survey[l] | | | | | | | |
| SDSS-V[g] | 2,000 | 0.37–1.00 | 500 | i < 20 | 7.0 | All sky | 2021– |
| | 22,000 | 1.51–1.70 | 300 | H < 13.4 | | | |
| DESI[h] | ~4,000 | 0.36–0.98 | 5,000 | z < 21.5 | 40.0 | Northern sky | 2021– |
| WEAVE[i] | 5,000 | 0.37–0.95 | 1,000 | G < 19 | 15.0 | Northern sky | – |
| | 20,000 | 0.41–0.46 | | | | | |
| | | 0.60–0.68 | | | | | |
| 4MOST[j] | 5,000 | 0.40–0.90 | 800 | r ≤ 22 | 10.0 | Southern sky | 2023– |
| | 20,000 | | 1,600 | V ≤ 16 | | | |
| MOONS[k] | >4,100 | 0.65–1.80 | 1,000 | H < 24 | 2.0 | Southern sky | 2022– |
| | 9,200 | 0.76–0.89 | | H < 18.5 | | | |
| | 18,300 | 1.52–1.64 | | | | | |

[a]LAMOST:[1] http://www.lamost.org/public/?locale=en.
[b]RAVE:[13] The Radial Velocity Experiment survey, https://www.rave-survey.org.
[c]SEGUE:[14] The Sloan Extension for Galactic Understanding and Exploration survey, https://www.sdss.org/surveys/segue/.
[d]APOGEE:[15] The Apache Point Observatory Galactic Evolution Experiment survey, https://www.sdss.org/surveys/apogee/.
[e]GALAH:[16] The Galactic Archaeology with Hermes survey, https://www.galah-survey.org.
[f]Gaia-ESO:[17] The Gaia-European Southern Observatory survey, https://www.gaia-eso.eu.
[g]SDSS-V:[18] The Sloan Digital Sky Survey' fifth generation, https://www.sdss5.org.
[h]DESI:[19] The Dark Energy Spectroscopic Instrument survey, https://www.desi.lbl.gov.
[i]WEAVE:[20] The William Herschel Telescope Enhanced Area Velocity Explorer, https://www.ing.iac.es//confluence/display/WEAV/The+WEAVE+Project.
[j]4MOST:[21] The 4-m Multi-Object Spectroscopic Telescope survey, https://www.4most.eu/cms/.
[k]MOONS:[22] The Multi-object Optical and Near-IR Spectrograph survey, https://vltmoons.org.
[l]Not all planned modes are listed.

**Binaries.** Repeated measurements of LAMOST RVs allow detection[110] and characterization of large samples of binary stars, in particular eclipsing binaries. Investigations of binary systems open the floodgates to a wealth of astrophysical knowledge. For instance, the combination of RV and photometry allows binary star orbital solutions to be comprehensively derived, which is crucial to constrain the theory of stellar evolution (eg, the exemplary case of systems containing a white dwarf and cool subdwarf[111]). In addition, binaries allow a more detailed analysis of stellar activity;[112] however, most binary systems are spatially unresolved and challenging to identify with a high degree of completeness. The synergy between Gaia and LAMOST provides an efficient way of identifying binary systems with high mass ratios via the difference between their geometric parallax and spectrophotometric parallax.[57]

While direct methods to identify binaries are usually biased to certain types of binaries[113] and limited to the solar neighborhood, large-scale surveys such as LA-

MOST enable statistical studies of binary fractions and properties. For example, by modeling variations in RVs from multiple LAMOST observations, binary fractions are found to be larger for metal-poor and hot stars,[114] whereas this conclusion is at odds for wide binaries.[115,116] Modeling color offsets with respect to the metallicity-dependent stellar locus, which is sensitive to neither the period nor mass-ratio distributions of binaries, Yuan et al. (2015)[117] provided a model-free estimate of the binary fraction for field FGK stars and found that the Galactic halo contains a larger fraction of binaries than the Galactic disk. Using the same technique, Niu et al. (2021)[118] found different effects of the chemical abundances on binary fractions for thin- and thick-disk stars, which are likely related to their distinct formation histories. By modeling magnitude offsets with respect to single stars, Liu (2019)[119] reported clear evidence of dynamic processes within solar-type field binary stars, which tends to destroy binaries with smaller primary mass, smaller mass-ratio, and wider separation in star clusters.

**Hypervelocity stars.** The huge number of LAMOST stellar spectra available provide a great opportunity to discover hypervelocity stars. Following the first discovery of a hypervelocity star using LAMOST data in 2014,[120] tens of hypervelocity stars or candidates have been identified.[121–126] Detailed analyses of this sample will be advantageous to understanding the diversity and origin of hypervelocity stars.

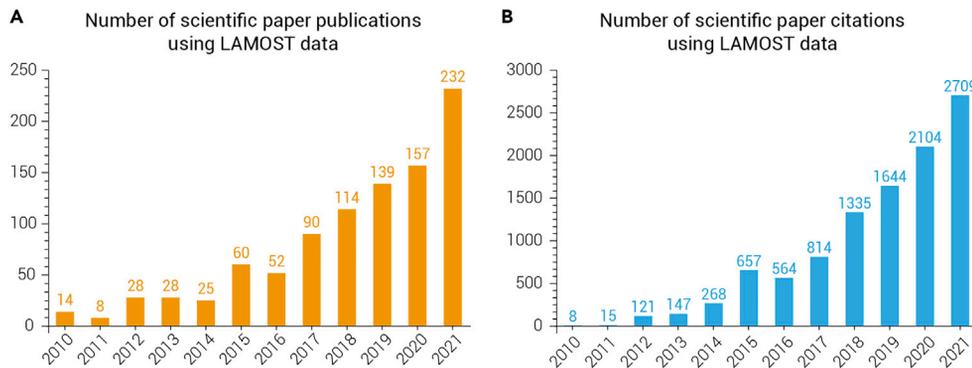

**Figure 4. Information of scientific papers using LAMOST data** Number of scientific papers published using LAMOST data (A) and the citation numbers to these works (B) up to the end of 2021. Data are from the Web of Science.

### Reshape the understanding of the MW

Constraining the formation and evolution of the galaxies is among the most challenging tasks facing modern astrophysics. Our own galaxy, the MW, provides a unique opportunity to study a galaxy in exquisite detail. The accomplishment of this goal relies on large surveys that characterize the position, motion, chemical compositions, and ages of millions of stars.[127] This was the key scientific goal of LAMOST as one of the most efficient spectroscopic survey telescopes in the world.[3,4]

***MW disk.*** LAMOST has conducted the first deep, spatially contiguous spectroscopic survey of the anticenter disk using a simple but well-defined target selection function.[5,6,128] The resulting data products, especially when combining with Gaia and other surveys, are improving our knowledge about the MW disk.

First, LAMOST data enable us to determine and renew some fundamental parameters of the MW disk: (1) new determinations of the velocity of the Sun[129] and confirmation that the value of $V_Z$ is larger than previous results.[130] (2) the accurate rotation curve of the MW disk;[131] (3) local dark matter density;[132,133] and (4) the size of the MW disk is found to be significantly larger than previously known[134] (Figure 7).

Second, LAMOST has yielded extensive new results about stellar populations, structures, and chemodynamic evolution of the MW disk. Using LAMOST data, a 3-dimensional stellar-mass density distribution to ~4 kpc from the Sun was precisely estimated for the first time.[47] The stellar mass was found to be distributed non-smoothly but highly structured throughout the MW disk. In particular, the stellar-mass density at a radius of ~8 kpc is significantly higher than previous estimates[135] by approximately 0.016 $M_\odot$/pc$^3$.[47] The disk flaring phenomenon was clearly revealed by LAMOST[47,136,137] (Figure 7), which was found to be a ubiquitous phenomenon for all monoage populations. Complex features in the stellar distribution throughout the outer disk revealed by LAMOST[47,138] reflect a significantly perturbed disk (Figure 7) evolution history.[139,140]

It is believed that the stellar disk has undergone significant external and internal perturbations.[141,142] Gaia data have led to discoveries of numerous fine structures in the stellar disk—for example, the "snail shell" feature in the $Z - V_Z$ phase-space[143] and the diagonal ridge patterns in $R - V_\varphi$ phase-space;[143,144] however, the scientific ability of the Gaia data alone is limited due to its lack of stellar RVs. The synergy between LAMOST and Gaia, as well as other surveys, has led to extensive studies into the disk velocity field.[49–51,141,145–151] For example, Tian et al. (2018)[145] found that the snail shell feature may be caused by perturbation during the last 500 million years. Wang et al. (2019)[146] found that the snail shell feature becomes more relaxed with increasing radius, consistent with predictions that the perturbation was caused by an external intruder. Wang et al. (2020)[50] studied the $R - V_\varphi$ diagonal ridge pattern of stellar populations with different ages, and found evidence for two types of dynamic origins. The stellar age-velocity dispersion relation was also studied extensively with LAMOST data.[152–154] Wu et al. (2021)[51] mapped the relation between kinematics, age, and metallicity, and found that stellar migrators in the solar nearby were not only from the inner disk but also from the outer disk.

Conducting a pioneering study of the radial and vertical metallicity gradients for monoage disk stellar populations from LAMOST, Xiang et al. (2015)[46] revealed that the disk assembly had experienced two phases that transited at 8–11 Gyr ago. Wang et al. (2019)[48] extended these studies to include full distribution function of the metallicity and $\alpha$-abundance, revealing a complex disk formation history that consists of both "inside-out" and "upside-down" processes. Huang et al. (2015)[155] measured the metallicity gradients for a significant volume of the MW disk and found that the outer disk may have experienced a different enrichment history compared to that at the solar nearby. Combining data from LAMOST and Gaia, Vickers et al. (2021)[156] studied the variation of disk radial metallicity gradients with stellar age and found clear evidence of radial migration. The disk metallicity gradients were also studied using LAMOST metallicities of star clusters.[157] By investigating the metallicity separation of action-angle pairs using a large sample of LAMOST dwarf stars, Coronado et al. (2020)[158] characterized the dissolution of stellar birth associations into the field.

***MW halo.*** During the past 2 decades, our knowledge of the Galactic halo has increased greatly, revealing a halo with large overdensities and many narrow streams.[159,160] Gaia DR2 has driven a further revolution with the unexpected discovery that the nearby halo is dominated by debris from a single accretion event with Gaia-Enceladus[161,162] (see Helmi, 2020[163] for the most recent review).

LAMOST has increased the number and reliability of identified stellar streams. Approximately 2,000 halo K-giants have been identified in LAMOST DR5, which belong to more than 40 groups, including a number of known substructures as well as new ones.[164] This first large sample in 6-dimensional phase-space presents the more comprehensive view yet obtained for the MW substructure.

The Sgr stream, the remnants of the Sgr dwarf spheroidal (dSph) galaxy, is the most prominent tidal stream around the MW[165] (Figure 7). It can be traced more than 360° across the sky[166] and provides a unique window to explore the formation history of the MW.[167] Using data from LAMOST and other surveys (eg, SDSS, Gaia), a variety of types of stars has been used to trace the Sgr tidal debris,[168–171] revealing details about the structure of the stream and providing stronger constraints on the complex star formation history of the Sgr dSph.

After stellar streams spread throughout space, their identification could only be made through observing structures in kinematics (e.g., moving groups[172] [MGs]). As such, detection becomes difficult, requiring both large sample sizes and effective methods. Based on the LAMOST data, seven new MGs have been identified,[142,173] comprising half of the known halo MGs. Follow-up studies enable us to reveal their origins. For example, high-precision abundances of MG LAMOST-N1 indicate that its progenitor may be a relatively large and early accreted dwarf galaxy.[174]

When stellar streams completely lose their structures in space and kinematics, searching for chemical imprints inherited from parent galaxies is the only way to identify them. LAMOST has enabled the first systematic search of halo stars deficient in $\alpha$-abundances (typical in dwarf galaxies[175]). More than 90 $\alpha$-deficient ([$\alpha$/Fe] < 0.0) halo stars have been discovered using LAMOST;[176,177] in other words, twice the size of the previously existing sample. Very recently, the discovery of an $\alpha$-deficient and neutron-capture element-enhanced halo star have been reported,[177] providing key evidence about its origin from a neutron star merger in a dwarf galaxy.

Using the LAMOST K-giant sample, the halo has been studied in detail. Its number density distribution clearly shows an oblate inner halo and a nearly spherical outer halo,[178] confirming its rotational velocity distribution.[179] These results also reveal the signal of the interaction between halo and disk, which possibly represents the mechanism that shapes the halo. LAMOST data enabled the first measurement of 3-dimensional velocity dispersion and anisotropy profiles out to 100 kpc from the Galactic center.[180,181] It was found that halo stars move along radially dominated orbits in the MW, showing a chemodynamic trend—in other words, metal-rich stars are on more radially dominated orbits than metal-poor stars,

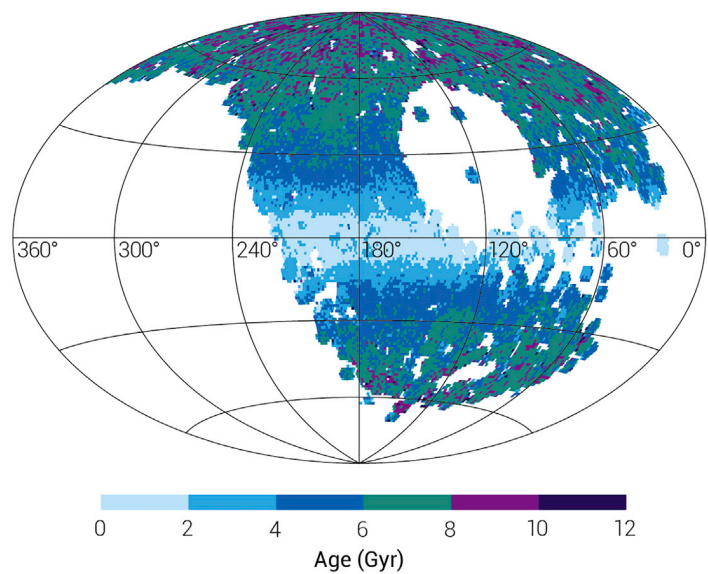

**Figure 5. Spatial variation of stellar age across our Galaxy in the Galactic coordinates (l,b)** The image center is the Galactic anticenter ($l = 180°$, $b = 0°$). The figure was adapted from Figure 17 of Xiang et al. (2017).[41]

which is likely a remnant of the ancient past merge of the Gaia-Enceladus Sausage.[161]

LAMOST also provides an opportunity to explore the evolution of the MW through very and extremely metal-poor (VMP/EMP, [Fe/H] < −2.0/−3.0) stars. Regarded as records of the early chemical history of the MW,[182] these objects have been extensively studied through large-scale surveys and follow-up observations.[183–185] Regrettably, the current sample of VMP/EMP stars is still limited and requires expanding. The most advanced search for VMP stars has been initiated[186] with LAMOST and has resulted in the largest bright VMP catalog, which includes more than 10,000 VMP stars and 670 EMP stars.[187] This catalog doubles the number of known EMP stars and provides crucial constraints on the evolution of the MW. Follow-up studies on this sample have opened new windows on Galactic history (e.g., the assembly of the oldest components of the MW).[188]

**MW dust.** As an important component of the MW, studies of the distribution and properties of dust in the Galaxy not only provide vital clues on Galactic structure and the formation and evolution of dust grains but also are crucial for the precise dereddening of astronomical targets. LAMOST data have made it possible to trace dust distributions and properties along millions of sight lines in unprecedented detail by combining multiband photometry with the star-pair technique[189] or the blue-edge method.[190] Li et al. (2018)[191] conducted accurate modeling of the MW dust component using reddening and distances from the LSS-GAC DR2. The result shows that the overall distribution of interstellar dust in the MW can be clearly described as a disk-like structure with a scale length of approximately 10,000 light-years and a scale height of approximately 330 light-years. Compared to the stellar disks of the MW, the dust disk is stretched more on the radial scale and thinner in the vertical. Chen et al. (2019)[192] constructed 3-dimensional interstellar dust-reddening maps of the Galactic plane. A near-UV extinction map of approximately one-third of the sky at high Galactic latitude has also been delineated.[193] Reddening distances of hundreds of molecular clouds,[194] including those at high Galactic latitudes,[195] and tens of SN remnants[196] have also been obtained. By careful subtraction of a 2-dimensional foreground dust-reddening map from that of Schlegel et al. (1998)[197] toward the M31 and M33 regions, Zhang and Yuan (2020)[198] detected dust disks extending to approximately 2.5 times the optical radii of M31 and M33, as well as a large amount of dust in the M31 halo out to 100 kpc. Reddening laws in the UV[195,199] and in SN remnants[196] have also been well studied, as have the intrinsic color indices of early-type dwarf stars.[200]

**Synergy between MW sciences and other surveys.** In addition to large-scale spectroscopic surveys such as LAMOST, wide-field photometric surveys are essential for mapping the stellar populations, structure, and chemistry of the Galaxy. Nonetheless, photometric estimates of stellar parameters are usually limited by systematic errors in photometric calibration. The availability of millions of high-quality stellar spectra and high-precision stellar parameters from the LAMOST survey permits the application of the stellar color regression (SCR) method[201–203] to calibrate photometric surveys to a precision of a few millimag. For example, Niu et al. (2021)[204,205] applied the SCR method to Gaia data, providing color corrections with millimag precision. The availability of millions of RVs from LAMOST also makes it possible to perform precise wavelength calibrations of data from the Chinese Space Station telescope slitless spectroscopic survey using stellar absorption lines;[206] this solves one of the most challenging problems in wide-field slitless spectroscopy.

### Probe the properties of planets beyond the Solar System

Much progress has been made recently in exoplanet science thanks to the Kepler mission, which has discovered thousands of transiting planets. Spectroscopic observations of planet hosts are crucial to characterize planets (i.e., to determine the radius of a planet, one needs to know the radius of the host star), study their statistics, and reveal connections between planet formation and evolution and stellar properties and environments. The LAMOST-Kepler project[7,91,93] has gathered the largest spectroscopic sample of Kepler targets, offering a uniquely large and homogeneous sample for Kepler planet statistics, as demonstrated by Dong et al. (2014)[207] and many subsequent works by numerous groups. Here, we review three exoplanet studies using LAMOST-Kepler data: revealing exoplanets' orbital patterns,[208] identifying a new planet population,[209] and determining the frequency of Kepler-like planetary systems.[210]

The orbits of planets in the Solar System are almost circular and coplanar, but hundreds of giant planets found by RV surveys have unusually eccentric orbits (average eccentricity $\bar{e} \sim 0.3$), presenting a great puzzle—is our Solar System special in this regard? Using LAMOST data and applying transit duration statistics, Xie et al. (2016)[208] measured the eccentricity distribution of Kepler planets, finding that single-transiting planets are commonly on eccentric orbits ($\bar{e} \sim 0.3$), whereas multiples are on nearly circular ($\bar{e} \sim 0.04$) and coplanar orbits. This eccentricity dichotomy was later confirmed by studies using asteroseismology[211] and Keck spectroscopic data.[212] It has important implications for understanding the dynamic evolution of exoplanets.[213] Xie et al. (2016)[208] also showed that the Kepler multiples and Solar System objects follow a common relation in their orbital patterns (Figure 8A), suggesting that the Solar System is probably not as atypical as once thought.

"Hot Jupiters" are Jovian planets with orbital periods p < 10 days, and their formation mechanisms are actively debated.[214] Recently, Dong et al. (2018)[209] identified a new population of close-in planets dubbed "Hoptunes" (Figure 8B), which have radii $R_p \sim 2 − 6 R_\oplus$ and share several key similarities with hot Jupiters. Both populations have preferentially metal-rich hosts and frequencies of ∼1% that have similar dependence with host [Fe/H]. Also, like hot Jupiters, Hoptunes tend to exist in systems with single-transiting planets. This empirical "kinship" implies that they likely share common migration and formation processes, providing new clues to the formation and evolution of close-in giant planets.[214]

A basic question addressed by Kepler planet statistics is the fraction of stellar hosting planetary systems detectable by Kepler. This requires knowing the average number of not only planets per star but also planets per Kepler planetary system, which degenerates with mutual inclination distributions when using only Kepler transit data.[215,216] Zhu et al. (2018)[210] broke this degeneracy by analyzing a homogeneous Kepler sample characterized by LAMOST combined with transit timing variations statistics, finding that systems with fewer transiting planets have higher mutual inclinations. They also found that approximately one-third of Sun-like stars have Kepler-like planetary systems ($R_p > R_\oplus$ and p < 400 days), and each system has on average 3 planets. These statistics are important for understanding the formation efficiency and dynamic evolution of such planetary systems.

### Reveal the hidden secrets of the deep Universe

In recent years, much exciting and significant progress has been made in the field of black holes (BHs), from the detection of gravitational waves to the first photograph of a BH. BHs can be divided into three classes according to their mass: stellar-mass BH (<100 $M_\odot$), intermediate mass black hole, and supermassive BH (>$10^6 M_\odot$, SMBH).

**Stellar-mass black holes.** Most identified Galactic stellar-mass BHs (∼20) were originally identified using X-rays, which are emitted from the gas that

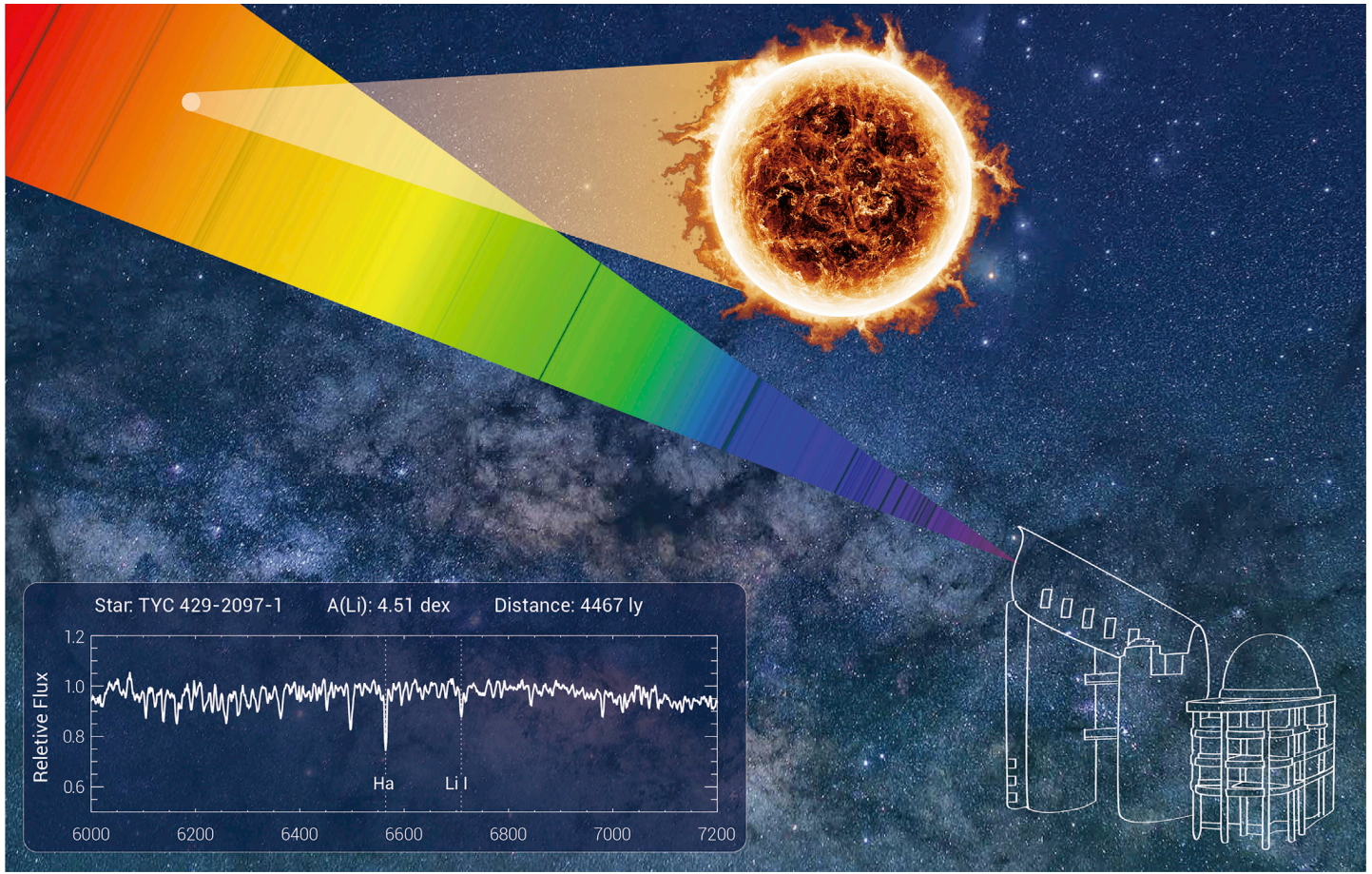

Star: TYC 429-2097-1    A(Li): 4.51 dex    Distance: 4467 ly

Relative Flux

Hα    Li I

**Figure 6. Artist's depiction of the most Li-rich giant TYC429-2097-1 and its spectrum (bottom left) taken by LAMOST** This star was discovered by the LAMOST-LRS. The Li I line in the subframe is clearly visible from R ~1,800 spectrum. The spectrum is from Yan et al. (2018).[67]

accretes onto the BH from the companion star. However, theories predict that the vast majority of star-BH binaries do not emit X-rays. These X-ray quiescent systems can be identified using RV measurements of the motion of the companion star.

LAMOST provides us with a great opportunity to successfully use this technique to discover BHs. From RV monitoring campaigns aiming to study spectroscopic binaries, Liu et al.[217] found that a B-type star, called LB-1, exhibits periodic (~79 days) RV variation, along with a strong, broad Hα emission line moving in antiphase. Combined with observations by the Gran Telescopio Canarias and Keck spectral observations, Liu et al.[217] reported one massive stellar-mass BH in the wide binary system (Figure 9). Furthermore, by conducting a high-resolution, phase-resolved spectroscopic study using the Calar Alto 3.5 m telescope, Liu et al.[218] found that the Pa $\beta$ and Pa $\gamma$ lines of LB-1 fit closely to a standard double-peaked disk profile, and derived a primary mass that is 4–8 times higher than the secondary mass. It is argued that LB-1 contains either a normal stellar-mass BH (~5–20 $M_\odot$)[219] or a Be star primary, rather than a BH, as well as a hot subdwarf companion.[220] The nature of LB-1 is still under debate, and further astrometric *Gaia* data may be helpful in distinguishing different scenarios.

By using LAMOST and photometric data (i.e., All-Sky Automated Survey for SN [ASAS-SN]), Gu et al.[221] and Zheng et al.[222] proposed a method to search for binaries containing a giant star and a compact object and presented a sample of candidates. In addition, by convolving the visibility of BH binaries with LAMOST detection sensitivity, Yi et al.[223] predicted that more than 400 BH binary candidates can be found with the non-time-domain LRS and that approximately 50–350 candidates can be detected with current time-domain MRS.

**SMBHs and galaxies.** Most large galaxies contain SMBHs, with masses ranging from millions to billions of solar masses in their galactic nuclei. When gas around the galactic center falls toward these SMBHs, an accretion disk forms, and energy is released in the form of electromagnetic radiation having almost all wavelengths. These active Galactic nuclei (AGN) may have luminosities as high as 10,000 times that of their host galaxies, called quasars.

The LAMOST quasar survey has identified more than 40,000 quasars, with the most distant having a redshift of z≈ 5.[224–226] LAMOST data has helped to investigate quasar spectral variability and discover unusual quasars. One such example is the changing-look (CL) AGNs, featuring the appearance and disappearance of broad Balmer emission lines within only a few years. CL AGNs are important for understanding the physical mechanism behind this type of transition and thus the evolution of AGNs. Yang et al.[227] found 21 new CL AGNs, 10 of which were discovered by LAMOST. By investigating their optical and midinfrared variability, the authors confirmed a bluer-when-brighter trend in the optical, but also found a redder midinfrared WISE color W1−W2 when brighter, possibly due to a strong contribution from the AGN dust torus when the AGNs turn on.

LAMOST also provides an opportunity to construct large samples of galaxies. By combining SDSS, LAMOST, and Galaxy And Mass Assembly data, Feng et al.[228] constructed the largest galaxy pair sample to date and provided the first observational evidence of galaxy merging timescales (1–2 Gyr). By using LAMOST data, Napolitano et al.[229] presented estimates of the central velocity dispersion of ~86,000 galaxies. The derived mass-$\sigma$ relation from the LAMOST data for both early-type and late-type galaxies are consistent with previous analyses. This implies that LAMOST spectra are suitable for studying galaxy kinematics and are invaluable for studying the structure and formation of galaxies and determining their central dark matter content.

## SCIENTIFIC PERSPECTIVE
### Future observations
Starting in 2018, the LAMOST-MRS will last for 5 years and the plan is to obtain millions of medium-resolution spectra for objects brighter than G = 15 mag. The plan is also to provide 60-epoch observations for ~200,000 stars. The scheduled survey aims to address a series of cutting-edge research topics, including time-domain astronomy, stellar physics, Galactic archaeology, emission nebulae, exoplanets and host stars, and compact objects. Furthermore, the LAMOST-MRS data will readily fit into the bulk data collected by the entire astronomical community, providing unique information that is complementary to that obtained from ongoing and planned spaceborne telescopes or ground-based survey projects.

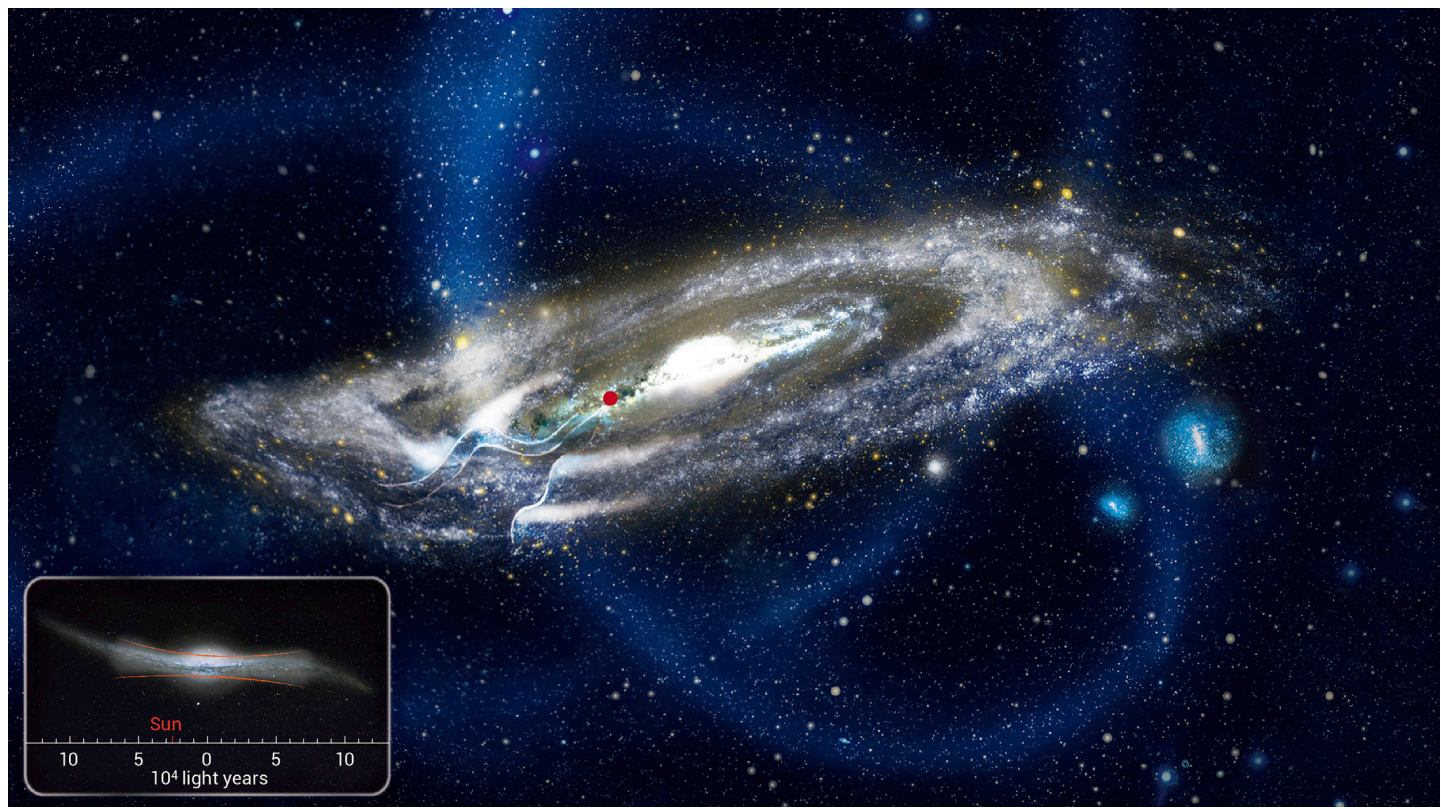

**Figure 7. Examples of how our understanding of the Milky Way (MW) has been improved by LAMOST data** It has been realized that the Galactic disk is not flat; rather, it has been warped and perturbated during the evolutionary history of the MW. Mergers with many smaller galaxies is believed to be a key mechanism that has forged the Galactic disk we see today. Smaller galaxies caught by the gravitational potential of the MW were stripped off, and eventually became tidal streams distributed around our Galaxy, which can be clearly seen from telescopes such as LAMOST. The light blue stream refers to one of the historical remains of such events, the remnants of the Sagittarius dwarf spheroidal galaxy. Our Sun is indicated with a red dot in the main frame. The subframe shows the perceived size of the MW, which has been enlarged from ~50,000 light-years to ~62,000 light-years, and then to ~100,000 light-years by LAMOST. The MW disk flaring phenomenon was also clearly revealed by LAMOST, as indicated by the red curve in the subframe. This figure is driven by scientific data but optimized for visualization.

### Perspectives for stellar physics

***Physical parameters of stars.*** The accurate characterization of stellar physical properties is essential to almost all topics addressed by LAMOST. A number of crucial parameters have been derived from the LAMOST-LRS data, including atmospheric parameters, RVs, the chemical compositions for a dozen elements, and even masses and ages. As observations continue, such information will be presented for millions more stars; most important, they will cover higher dimensions (e.g., abundances for more species, RV variations) with better precision along with the increased resolution and developed analysis techniques. These fundamental parameters will provide the most basic properties of stars in the MW.

***Stars with peculiar spectral signatures.*** The continuous expansion of the LAMOST spectral database assists in the search for and the investigation of stars with various spectral signatures. The data in the LAMOST-LRS have already resulted in the discovery of a series of such objects. With the LAMOST-MRS, we can expand our research to cover a much wider range of topics and to make advances in a series of important questions. For example, the origin of heavy elements, especially elements synthesized through the *r*-process, is still under debate. Data from the LAMOST-MRS will enable us to search for the *r*-process elements-enhanced stars, and will help us to understand their true origins.

***Variable stars.*** Combining time-domain spectra and high-quality photometry, variations in RV and atmospheric parameters can be used to constrain the dynamic processes of high-amplitude pulsating stars and orbital solutions to eclipsing binaries, as well as to study variations of stellar activity strength. It is foreseen that examining many such targets will provide a statistically meaningful view of this field.

***Binary and multiple stars.*** With approximately five times more accurate RVs compared to those from low-resolution spectra, the LAMOST-MRS will help us to discover more binaries under various evolutionary stages, obtain their orbital parameters, and further constrain their properties. In addition, massive multiepoch observations will enable statistically significant investigations on the binary fraction and orbital properties for a large sample of stars, as well as their variations with different stellar parameters.

***Star-forming regions.*** A number of important questions regarding star formation have yet to be answered. The LAMOST-MRS will cover several star-forming regions. As well as investigating the binary fraction and possible dynamical processing of young stellar populations by multiepoch observations, the LAMOST-MRS data will also provide an insight into the protoplanetary disks that may surround the newly formed low-mass stars. It will also make it possible to determine the Li evolution of young stars with precise Li abundances measured through the LAMOST-MRS data.

***Emission line nebulae.*** Emission line nebulae connect the evolution of stars with that of the MW in the sense that these objects are either remnants of dead stars (e.g., planetary nebulae; SN remnants) or associated with newly born stars. Studies of emission line nebulae cover a broad range of topics, from stellar physics to the Galactic evolution. The LAMOST-MRS data will provide a valueless database since the spectra cover a number of key emission lines that can be used to characterize the physical properties of the nebulae (eg, H$\alpha$, [N II], and [S II]).

### Perspectives for MW studies

***Chemical tagging.*** Chemical tagging uses massive stellar abundances seen in the present Galaxy to reveal the history of the Galaxy by identifying its star-forming events. In general, the abundance of each element is treated as a chemical dimension to characterize a particular set of stars. Thus, the information of abundances covering a broad number of elements for millions of stars is ideal for chemical tagging. The LAMOST-MRS data enable us to analyze more elemental species with higher accuracy. In addition, combining these data with RVs and Gaia astrometry will provide a more comprehensive view of the MW.

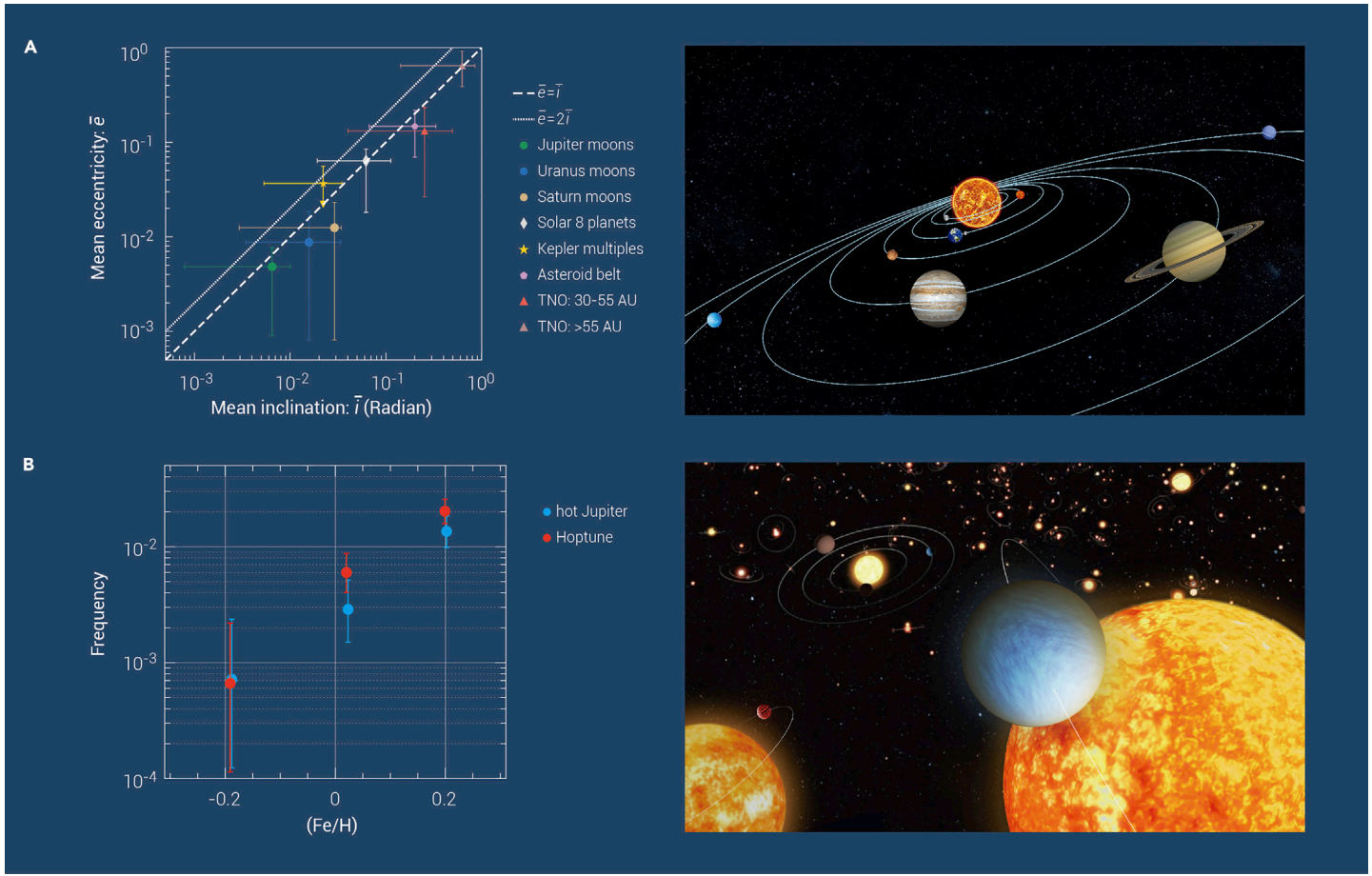

**Figure 8. Examples of planetary science done by LAMOST data** (A) The comparisons between Kepler multiple transiting systems and objects in the Solar System. It was found that they follow a common relation[198] between mean eccentricities and mutual inclinations: $\bar{e} \sim (1-2) \times \bar{\iota}$. The error bars show the 68% confidence intervals of orbital eccentricity and inclination distributions. (B) The comparisons between hot Jupiter and Hoptune; the latter is a recently identified exoplanet population.[209] A Hoptune is shown with its host star in the artist's depiction (credit: Jin Ma, Beijing Planetarium) at bottom right.

**Constraining the mass distribution of the MW dark halo.** Through 6-dimensional phase-space parameters of different types of stars, including K-giants and blue horizontal-branch (BHB) stars, the rotation and anisotropy of velocities can be obtained by adopting dynamical models that can simultaneously fit multiple populations of stars. Combined with the southern sky coverage by SDSS-V, a much larger sky area will be covered than ever before. More reliable measurements will undoubtedly constrain the total mass of the MW more precisely.

**Chemokinematic analysis of MW merging history.** A new halo star sample will be obtained with 6-dimensional phase-space parameters, metallicities, and abundances of key elements, which will allow the identification of new stellar streams and substructures, and stars with low α-abundances. Compared with high-precision MW dynamic simulation, these merging relics will help constrain the properties of their progenitor stellar systems. Questions such as whether merging processes have dominated the formation of the MW halo and which types of dwarf galaxies have been accreted in the past can be better answered; this will (largely) reproduce the merging history of the MW.

**Tracing the early evolution of the MW halo.** Larger samples of metal-poor halo stars with broad coverage in metallicities will better represent various halo components and can be used to systematically investigate the metallicity distribution function (MDF). Compared with model predictions, the low-metallicity end of the observed MDF will provide essential constraints on the enrichment history of the early MW. The identification of peculiar chemical sequences within this sample will provide an abundance pattern from which we can constrain basic parameters, such as the stellar masses of their progenitors. Kinematics shall be obtained for the majority, allowing the origin of different components of the MW halo to be revealed.

## Perspectives for planetary science

Our knowledge of exoplanets has been expanding from the solar neighborhood to a wider range of the MW. A Galactic census of exoplanets and investigation of their dependence on planet host properties and Galactic environments can shed light on the mechanism of planet formation and evolution. LAMOST spectra combined with *Gaia* astrometry will play a crucial role in studying exoplanets in the Galactic context. In addition, rich information on the elemental abundances provided by the LAMOST-MRS and time-series RVs will expand the means of understanding the stellar environments of planet formation and evolution.

## Perspective for compact object research

For the field of compact objects, the discovery of LB-1 suggests that future similar campaigns will probe a quiescent BH population different from the X-ray-bright population, which may rewrite our current understanding that only approximately five stellar-mass BHs were discovered with RV monitoring.[217,230−233] The LAMOST-MRS time-domain data would be extremely helpful in discovering X-ray-quiescent BHs. Together with the binary stellar-mass BHs discovered by the Laser Interferometer Gravitational-Wave Observatory and Virgo, complete mass distribution of BHs will be constructed and should help us to understand the evolutionary history of massive stars and the formation of BHs.

## SUMMARY

LAMOST observations have fundamentally modified our understanding of the Universe. Designed in the 1990s and beginning its survey in 2011, LAMOST is one of the most powerful telescopes in the world. LAMOST is innovatively designed with a uniquely large aperture and a wide FOV. LAMOST has released ~17 million spectra from various celestial objects. Using those data, we have deepened our knowledge of the Universe. LAMOST has changed the astrophysical viewpoint on a number of cutting-edge fields, including stars, the MW, exoplanets, and BHs. With the launch of the LAMOST-MRS, LAMOST has expanded its data application to a much wider research field. Astronomical research is entering a new era featuring both massive data quantities and higher data accuracy with the help of LAMOST and other telescopes.

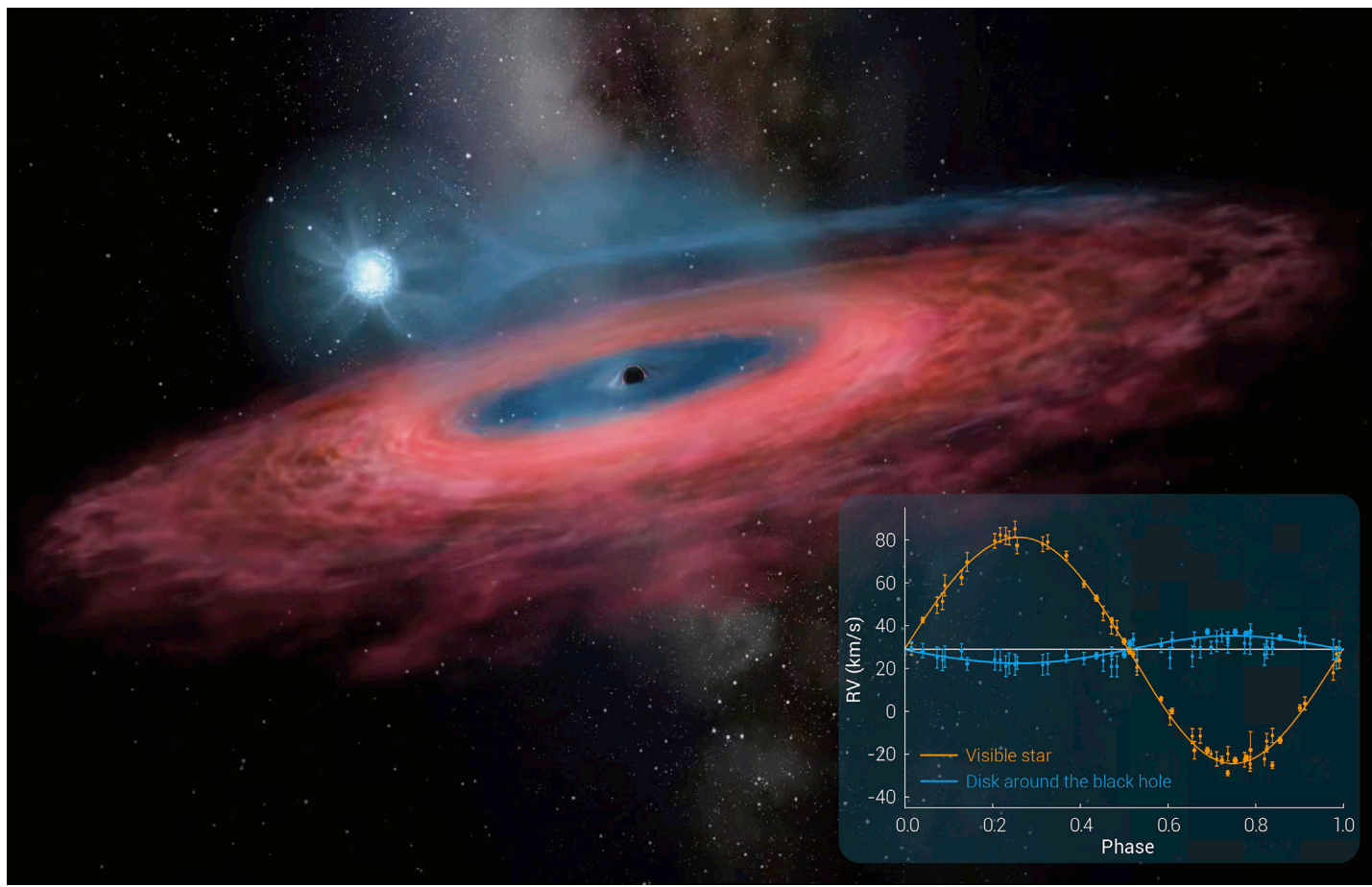

**Figure 9. Artist's impression of LB-1 showing the accretion of gas onto a stellar BH from its blue companion star through a truncated accretion disk** The subpanel shows the folded RV curves and binary orbital fits of the visible star (yellow curve with data points) and the dark primary (blue curve with data points). The observed data are from Liu et al. (2019).[217]

## ACKNOWLEDGMENTS

This work is supported by the National Natural Science Foundation of China under grant nos. 11988101, 11973049, 11933004, 11890694, 12090040, 12090042, 12090043, 12090044, 11833002, 11833006, 12022304, 11835057, 11973052, 11633005, 12173007, 11933001, 11703035, U2031203, and U1531244, the National Key R&D Program of China under grant nos. 2019YFA0405500, 2019YFA0405502, 2019YFA0405503, 2019YFA0405504, 2016YFA0400804, and 2019YFA0405000, and the Strategic Priority Research Program of the Chinese Academy of Sciences, grant nos. XDB34020205 and XDB41000000. H. Yan, H.L., S.W., and Hailong Yuan acknowledge support from the Youth Innovation Promotion Association of the Chinese Academy of Sciences (nos. 2019060, Y202017, 2019057, and 2020060, respectively). H. Yan and H.L. are supported by the NAOC Nebula Talents Program. We greatly thank Dr. S.A. Bird for the language editing, and Dr. D. Wang, Dr. S. Li, Dr. Q. Gao, and C. Li. for providing the statistical materials of LAMOST. Guoshoujing Telescope (LAMOST) is a National Major Scientific Project built by the Chinese Academy of Sciences. Funding for the project has been provided by the National Development and Re-form Commission. LAMOST is operated and managed by the National Astronomical Observatories, Chinese Academy of Sciences.

## AUTHOR CONTRIBUTIONS

Y.Z. proposed and supervised the project with the help of H. Yan. H. Yan, H.L., S.W., W.Z., H. Yuan, M.X., Y.H., J.X., S.D., and Hailong Yuan prepared the manuscript with the help of S.B., Y.C., X.C., L.D., J.F., Z.H., J.H., G.L., C.L., J.L., X.L., A.L., J.S., X.W., H.Z., and G.Z. All of the authors discussed and helped to revise the manuscript.

## DECLARATION OF INTERESTS

The authors declare no competing interests.

## SUPPLEMENTAL INFORMATION

Supplemental information can be found online at https://doi.org/10.1016/j.xinn.2022.100224.

## LEAD CONTACT WEBSITE

http://www.lamost.org/public/.